\documentclass[aps,prd,amsmath,floats,floatfix, twocolumn,
superscriptaddress,nofootinbib,showpacs,longbibliography]{revtex4-1}
\usepackage{color}
\usepackage{calc}
\usepackage{amsmath,amssymb,graphicx}
\usepackage{amssymb,amsmath}
\usepackage{bm}
\usepackage{microtype}
\usepackage{booktabs}
\usepackage{times}
\usepackage{subfigure}
\usepackage[varg]{txfonts}
\usepackage[colorlinks, pdfborder={0 0 0}]{hyperref}
\definecolor{LinkColor}{rgb}{0.75, 0, 0}
\definecolor{CiteColor}{rgb}{0, 0.5, 0.5}
\definecolor{UrlColor}{rgb}{0, 0, 0.75}
\hypersetup{linkcolor=LinkColor}
\hypersetup{citecolor=CiteColor}
\hypersetup{urlcolor=UrlColor}
\usepackage[utf8]{inputenc}

\DeclareFontFamily{OT1}{pzc}{}
\DeclareFontShape{OT1}{pzc}{m}{it}{<-> s * [1.10] pzcmi7t}{}
\DeclareMathAlphabet{\mathpzc}{OT1}{pzc}{m}{it}

\newcommand{\h}{\mathpzc{h}}

\newcommand{\hlm}{\mathpzc{h}_{\ell m}}

\newcommand{\Ylm}{{Y}^{-2}_{\ell m}}

\newcommand{\blambda}{\bm{\lambda}}

\newcommand{\n}{\mathbf{n}}

\usepackage{blindtext}

\graphicspath{
	{figs/}
}

\begin{document}
	
	\title{Applying higher-modes consistency test on GW190814 : \\ 
	lessons on no-hair theorem, nature of the secondary compact object and waveform modeling}
	
	\newcommand{\UMassDMath}{\affiliation{Department of Mathematics,
			University of Massachusetts, Dartmouth, MA 02747, USA}}
	\newcommand{\UMassDPhy}{\affiliation{Department of Physics,
			University of Massachusetts, Dartmouth, MA 02747, USA}}
	\newcommand{\CSCVR}{\affiliation{Center for Scientific Computing and Visualization 	Research, University of Massachusetts, Dartmouth, MA 02747, USA}}
	\newcommand{\KITP}{\affiliation{Kavli Institute for Theoretical Physics, University of California, Santa Barbara, CA 93106, USA}} 
	
	\author{Tousif Islam}
	\email{tislam@umassd.edu}
	\CSCVR
	\UMassDPhy
	\UMassDMath
	\KITP
	
	\hypersetup{pdfauthor={Islam}}
	
	\date{\today}
	
%==========================================================================
\begin{abstract}
As one of the consequences of the black-hole ``no-hair'' theorem in general relativity (GR), the multipolar structure of the radiation (i.e. different spherical harmonic modes) from a merging quasi-circular binary black hole (BBH) is fully determined by the intrinsic parameters (i.e. the masses and spins of the companion black holes). 
In Refs. \cite{Islam:2019dmk,Dhanpal:2018ufk}, we have formulated an efficient test named `higher-modes consistency test' to check for the consistency of the observed gravitational-wave (GW) signal with the expected multipolar structure of radiation from BBHs in GR.
Detection of the high-mass-ratio merger of GW190814 enables the observation of spherical harmonic modes beyond the dominant $(\ell, m) = (2,\pm2)$ mode; thereby providing a unique opportunity to perform the `higher-modes consistency test'.
Using different state-of-art waveform models (\texttt{IMRPhenomXPHM}, \texttt{IMRPhenomXHM}, \texttt{IMRPhenomHM} and \texttt{SEONNRv4HM\_ROM}), we show that GW190814 strongly favors the ``no-hair'' hypothesis in GR over a hypothesis that assumes generic deviation from the multipolar structure of the radiation by a Bayes factor of $\rm log_{e}\mathcal{B}\sim8$.
We further investigate any potential systematic errors arising as a result of different waveform modeling choices as well as due to neglecting many higher order modes. 
We find that, if the waveform model includes only $(\ell, m) = (3,\pm3)$ mode in its higher harmonics, the same event may reject the `no-hair' hypothesis in GR when the detector sensitivity improves by a factor of $12.5$.
Our analysis, therefore, provides motivation to include as many higher order modes as possible in future waveform models.
\end{abstract}

\maketitle

%==========================================================================
%==========================================================================
%==========================================================================
\section{Introduction} 
\label{Sec:Introduction}
GW190814~\cite{LIGOScientific:2020zkf,LIGOScientific:2020ibl} is the highest mass ratio binary merger event detected by the LIGO~\cite{LIGOScientific:2014pky}-Virgo~\cite{VIRGO:2014yos} collaboration so far. The original LVC analysis constrains the mass ratio to be $q:=m_1/m_2\sim8.9$~\cite{LIGOScientific:2020zkf} (where $m_1$ and $m_2$ are the component masses for the binary with $m_1>m_2$). Observation of GW190814 is also unique in the sense that the secondary object has a mass of $2.6 M_{\odot}$ - which lies in the hypothesized `lower mass gap' \cite{Bailyn_1998,Farr:2010tu,Ozel:2012ax,Ozel:2010su} defined by the theoretically possible highest mass for a neutron star and the lowest possible mass for a black hole. Till date, the exact nature of the secondary is still unclear~\cite{Goncalves:2021pmr,Das:2021yny,Rocha:2021zos,Nathanail:2021tay,Biswas:2020xna,Bombaci:2020vgw,Cao:2020zxi,Nunes:2020cuz,Huang:2020cab,Dexheimer:2020rlp,Clesse:2020ghq,Tews:2020ylw,Zhang:2020zsc,Essick:2020ghc,Vattis:2020iuz,Broadhurst:2020cvm}. This prompted fresh looks in understanding possible formation channels for this binary~\cite{ArcaSedda:2021zmm,Lu:2020gfh,Liu:2020gif,Takhistov:2020vxs} and its implications in alternate theories of gravity~\cite{Wang:2021jfc,Astashenok:2020qds,Moffat:2020jic}.
Such asymmetric mass-ratio systems excite several higher order harmonics of the gravitational radiation. The official LVC analysis~\cite{LIGOScientific:2020zkf} have been able to find strong support for the $(\ell,m)=(3,\pm3)$ spherical harmonics modes in the signal apart from the dominant quadrupolar $(\ell,m)=(2,\pm2)$ modes.
GW190814 therefore provides an excellent opportunity to probe beyond the dominant multipole of the GW signal. This event is particularly suited for performing a specific kind of null tests of binary black holes in general relativity (GR) - which we call the `higher modes consistency' tests~\cite{Islam:2019dmk,Dhanpal:2018ufk}.

`Higher modes consistency' test is one of the null tests of GR or of binary black hole nature~\cite{LIGOScientific:2020tif,Krishnendu:2017shb,Krishnendu:2019tjp,Johnson-Mcdaniel:2018cdu,Kastha:2018bcr,Kastha:2019brk,Carullo:2018gah,Carullo:2019flw,
Carullo:2021dui,Ghosh:2021mrv,Haster:2020nxf,Asali:2020wup,Edelman:2020aqj,Psaltis:2020ctj,Bhagwat:2021kfa,Ghosh:2017gfp, Ghosh:2016qgn,Isi:2019asy,Isi:2019aib,Isi:2020tac,Bustillo:2020buq}. The test exploits a generalized version of `no-hair theorem' for binary black holes in GR. 
``No-hair'' theorem in GR states that a stationary black hole can be fully described solely by its mass, spin angular momentum and electric charge~\cite{Israel:1967wq,Israel:1967za,Carter:1971zc}. As astrophysical black-holes are unlikely to posses electric charge, a stationary black hole can be fully described by only its mass and spin. As a consequence of this theorem, the gravitational radiation from a perturbed black hole can also be fully described by its masses and spins. 
This implies that the frequencies and damping times of the quasi-normal modes~\cite{Vishveshwara:1970zz,Press:1971wr,Chandrasekhar:1975zza} of the gravitational radiation from a perturbed black hole are fully determined by these parameters. This allows us to construct a test of `no-hair theorem' where we estimate the masses and spins of the perturbed black hole using different quasi-normal modes and compare those estimates against each other to check for consistency. If `no-hair theorem' is correct, these estimates have to be consistent with each other; otherwise it will indicate a potential violation of the no-hair theorem~\cite{Dreyer:2003bv}. This exercise is popularly known as `black-holes spectroscopy'~\cite{Dreyer:2003bv,Berti:2005ys,Berti:2007zu,Gossan:2011ha,Berti:2016lat,Giesler:2019uxc,Cabero:2019zyt,Isi:2021iql,Isi:2019aib}. 

Similarly, the gravitational radiation from a merging binary black hole can be uniquely determined in GR by a small set of parameters: masses and spins of the black holes and orbital parameters. As most binaries are expected to be circularized by the time it enters the LIGO-Virgo sensitivity band, we can safely ignore the eccentricity parameters. This implies that different spherical harmonic modes of the radiation have to be consistent with the same values for this small set of parameters. Thus, the consistency between the estimated values from different modes of the observed signal provides a test similar to the `black-holes spectroscopy'. Inconsistencies between different spherical harmonic modes would point to departure from the multipolar structure of radiation as expected from a binary black hole merger in GR. Such a violation would indicate that either the signal is produced by non-binary black hole compact objects or the underlying theory of gravity is not GR or the signal might have been modified by astrophysical phenomena such as lensing of GW signal \cite{Takahashi:2003ix} or environmental effects \cite{Barausse:2007dy,Yunes:2011ws,Barausse:2014tra,Barausse:2014pra}. In Refs. \cite{Islam:2019dmk,Dhanpal:2018ufk}, we have developed two such tests (that we refer to as `higher modes consistency' tests) that can efficiently check for possible inconsistencies in different spherical harmonic modes of the observed GW signal by introducing generic deviation parameters for the masses in the higher order modes. Similar test~\cite{Capano:2020dix} have recently been applied to GW190412~\cite{LIGOScientific:2020stg} and GW190814 events to constrain deviation from GR. 

In this article, we attempt to perform the `higher modes consistency' test as proposed in Refs. \cite{Islam:2019dmk,Dhanpal:2018ufk} on GW190814 with a focus on testing the validity of `no-hair theorem' (or, in other words, the multi-polar structure of radiation as expected in GR) and quantify its evidence against scenarios favoring possible violations. We note that the efficiency of such test may also be dependent on the accuracy of models for gravitational radiation being used. Different gravitational waveform models employ different modeling strategies and have varying degree of accuracy when compared to numerical relativity (NR) data. These models mostly fall in three different categories: effective-one-body (EOB) waveform family \cite{bohe2017improved,cotesta2018enriching,cotesta2020frequency,pan2014inspiral,babak2017validating,cotesta2020frequency, pan2014inspiral,babak2017validating}, phenomenological (Phenom) waveform family \cite{husa2016frequency,khan2016frequency,london2018first,khan2019phenomenological,hannam2014simple,khan2020including}, and NR based surrogate waveforms \cite{varma2019surrogate,varma2019surrogate2,Islam:2021mha}. To understand how waveform systematic may affect the test accuracy, we use four different waveform models in our analysis. We also investigate the effects of the availability of different higher order modes. While numerous studies have focused in understanding possible biases in usual parameter inference of the detected GW signals due to not including higher order modes~\cite{Varma:2016dnf, bustillo2016impact, Capano:2013raa, Littenberg:2012uj,Bustillo:2016gid, Brown:2012nn, Varma:2014, Graff:2015bba, Harry:2017weg, chatziioannou2019properties, Shaik:2019dym,Islam:2021zee}, such exercises have rarely been taken up for null-tests of GR like the `higher modes consistency' test~\cite{Johnson-McDaniel:2021yge}. 

The rest of the paper is organized as follows. Section \ref{Sec:tgr_hm} presents a brief outline of the higher modes consistency test and Section \ref{Sec:setup} describes the data analysis framework including Bayesian inference. In Section \ref{Sec:results}, we perform the higher modes consistency test on GW190814 strain data using different waveform models and modes content. We then quantify the evidence for the `no-hair theorem' in GR by computing Bayes factors. In this section, we also investigate the effects of higher order harmonics by employing different combination of the higher modes. We show that including only select higher order modes may lead to false identification of a departure from the `no-hair theorem' if the detector sensitivity improves by almost an order of magnitude. Finally, in Section \ref{Sec:conclusion}, we summarize our results and discuss the implications of our findings.

%==========================================================================
%==========================================================================
%==========================================================================
\section{Higher modes consistency test} 
\label{Sec:tgr_hm}

\subsection{Gravitational radiation in GR}

Gravitational radiation from the coalescence of a binary black hole in GR is usually written as a linear combination of two independent polarizations (`plus' and `cross'):
\begin{eqnarray}
\h_{\rm GR}(t; t_c, \n, \blambda) = h_{+,\rm GR}(t; t_c, \n, \blambda) - i \, h_{\times,\rm GR}(t; t_c, \n, \blambda), 
\label{eq:polarizations}
\end{eqnarray}
where $t_c$ is time at coalescence. 
This radiation can further be decomposed into a basis of $-2$ spin-weighted spherical harmonics ${\Ylm}$ ~\cite{Newman:1966ub}:
\begin{eqnarray}
\h_{\rm GR}(t; t_c, \n, \blambda) = \frac{1}{d_L}\sum _{\ell=2}^{\infty} \sum _{m=-\ell}^{\ell} {\Ylm} (\n) \, {{\hlm}(t; t_c, {\blambda})}, 
\label{eq:spherical_harmonics}
\end{eqnarray}
The set of two angles $\n=\{\iota, \varphi_0\}$ denotes the direction of radiation in the source frame: $\iota$ is the inclination angle between the orbital angular momentum of the binary and line-of-sight to the observer and $\varphi_c$ is the azimuthal angle 
at coalescence. 
The vector $\blambda:=\{\mathcal{M}_c,q,\chi_1,\chi_2,\theta_1,\theta_2,\phi_{12},\phi_{jl}\}$ are the intrinsic parameters that describe the binary: the chirp mass $\mathcal{M}_c$, mass ratio $q$, dimensionless spin magnitudes $\chi_1$ and $\chi_2$, and four angles $\{\theta_1,\theta_2,\phi_{12},\phi_{jl}\}$ describing the spin orientation (cf. Appendix of \cite{Romero-Shaw:2020owr} for definitions of these angles).  Luminosity distance of the binary from the observer is $d_L$. So far, all GW detections of BBHs are consistent with signals emitted from quasicircular binaries. We, therefore, ignore eccentricity in our analysis.

Finally, the gravitational signal  $h(t)$ detected by interferometers is a linear combination of the plus and cross polarization of the radiation weighted by the antenna pattern functions of the GW detector $F_+$ and $F_\times$:
\begin{eqnarray}
h_{\rm GR}(t;\theta_{\rm GR}) & = & F_+(\alpha, \delta, \psi) \, h_{+,\rm GR}(t; t_c, \n, \blambda) \nonumber \\ 
& + &  F_{\times}(\alpha, \delta, \psi)\, {h}_{\times,\rm GR}(t; t_c, \n, \blambda), 	
\label{eq:det_response}
\end{eqnarray}
where right ascension $\alpha$ and declination $\delta$ are the sky localization angles and $\psi$ is the polarization angle.
Together, this set of 15 parameters $\theta_{\rm GR}:=\{t_c, \varphi_0, \iota, d_L, \alpha, \delta, \psi, \blambda\}$ describes a gravitational wave signal for binary black hole merger in GR.

\subsection{Formulation of the higher modes consistency test}
For quasi-circular binaries in GR, the set of \emph{intrinsic} parameters $\blambda$  i.e., the masses and spins of the two black holes uniquely determines each of the spherical harmonic modes, ${\h}_{lm}(t; t_c, \blambda)$.  
The consistency of different spherical harmonic modes of the radiation provides a unique test of the ``no-hair'' nature of binary black holes in GR. 
This involves estimating the intrinsic parameters of the binary from different spherical harmonic modes of the radiation and checking their consistency. 
Inconsistency between different spherical harmonic modes would indicate a departure from the multipolar structure of the radiation as expected from a binary black hole system in GR. 
This, however, requires that different spherical harmonics modes have sufficient signal-to-noise ratios so that they can be well resolved.

Alternatively, one can look for consistencies between the dominant $(\ell,m) = (2,\pm 2)$ mode of the gravitational radiation, and the sub-dominant modes.
Following ~\cite{Islam:2019dmk,Dhanpal:2018ufk}, we allow inconsistencies between the intrinsic parameters estimated from the dominant mode and the higher order modes by introducing a set of deviation parameters for masses in the higher modes:
\begin{equation}
\Delta \blambda := \{\Delta \mathcal{M}_{c}, \Delta q\}.
\end{equation}
Gravitational radiation then reads:
\begin{eqnarray}
\h_{\rm nonGR}(t; t_c, \n, \blambda, \Delta \blambda) & = &    \sum_{m = \pm2} Y^{-2}_{2m} (\n) {\h}_{2m}(t, \blambda) \nonumber \\ 
& + &  \sum _{\text{HM}} \Ylm (\n) \hlm(t, \blambda, \Delta \blambda).	
\label{eq:test}
\end{eqnarray}
Here, {HM} indicates sum over higher order modes i.e. all modes other than $(\ell,m) = (2,\pm 2$). In terms of the `plus' and `cross' polarizations this simply becomes:
\begin{eqnarray}
\h_{\rm nonGR}(t; t_c, \n, \blambda, \Delta \blambda) &=& h_{+,\rm nonGR}(t; t_c, \n, \blambda, \Delta \blambda)  \nonumber \\
&-& i \, h_{\times,\rm nonGR}(t; t_c, \n, \blambda, \Delta \blambda).
\label{eq:polarizations_modgr}
\end{eqnarray}
The test therefore have a set of 17 parameters to describe the signal: $\theta_{\rm nonGR} =\{\theta_{\rm GR},\Delta \blambda\}$.
For GW signals produced by binary black hole mergers in GR, these additional parameters will be consistent with zero: $\Delta \blambda =\{0,0\}$. If the signal is not consistent with that produced by a BBH in GR, $\Delta \blambda$ will have non-zero values.

\section{Data analysis framework} 
\label{Sec:setup}
We now provide an executive summary of the data analysis framework used to perform the higher modes consistency test on GW190814.
\subsection{Bayesian Inference}
The measured strain data in a GW detector, 
\begin{equation}
d(t)=h(t;\theta)+n(t),
\end{equation}
is assumed to be a sum of the true signal, $h(t;\theta)$, and random Gaussian and stationary noise, $n(t)$ with zero mean and a power spectral density (PSD), $S_n(f)$. Here, $\theta$ is the set of parameters that describes the BBH signal that is embedded in the data.
Given the time-series data $d(t)$ and a model for GW signal $H$ (i.e. whether the signal is consistent with binaries in GR or not), we use Bayes' theorem to compute the \textit{posterior probability distribution} (PDF) of the binary parameters,
\begin{equation}
p(\theta | d, H) = \frac{\pi(\theta | H) \mathcal{L}(d|\theta, H)}{\mathcal{Z}(d | H)},
\end{equation}
where $\pi(\theta | H)$ is the \textit{prior} astrophysical information of the probability distributions of BBH parameters $\theta$ and $\mathcal{L}(d|\theta, H)$ is the likelihood function describing how well each set of $\theta$ matches the assumptions of the data. 
$\mathcal{Z}(d|H)$ is called the model evidence or marginalized likelihood.
For binaries in GR, we replace $h(t;\theta)$ with $h(t;\theta_{\rm GR})$ (and $\theta$ with the 15-dimensional set of parameters $\theta_{\rm GR}$) whereas, while performing the higher modes consistency test, we use $h(t;\theta)=h(t;\theta_{\rm nonGR})$ (and $\theta=\{\theta_{\rm GR},\Delta \blambda\}$).
We also compute Bayes factors~\cite{Veitch:2009hd} in favor of the GR hypothesis
\begin{equation}
\mathcal{B}=\frac{\mathcal{Z_{\rm GR}}}{\mathcal{Z_{\rm nonGR}}},
\end{equation}
where $\mathcal{Z_{\rm GR}}$ and $\mathcal{Z_{\rm nonGR}}$ denote the evidence for a signal that is consistent with the multipolar structure as expected for a binary black hole in GR and signal that deviates from the expected behavior in GR respectively.
The Bayes factor quantifies how much more likely that the data is described by a signal consistent with the `no-hair' theorem in GR.

\subsection{Setup}
To compute the posterior probability distribution of BBH parameters $p(\theta | d, H)$, we use the Bayesian inference package \texttt{parallel-bilby} \cite{ashton2019bilby,smith2019expediting,Romero-Shaw:2020owr} with \texttt{dynesty} \cite{speagle2020dynesty} sampler. 
We obtain the GW190814 strain and PSD data for all three detectors (LIGO-Hanford, LIGO-Livingston and Virgo) from the Gravitational Wave Open Science Center~\cite{GWOSC_GW190814, LIGOScientific:2020zkf}. 
The PSDs for these data were computed through the on-source \texttt{BayesWave} method~\cite{Littenberg:2014oda,Cornish:2014kda,Chatziioannou:2019zvs}, and use the inferred median PSDs in our analysis following the same assumptions as in Ref.~\cite{LIGOScientific:2020zkf}.
Following the LVC analysis of the event, we start our data analysis at 20Hz for LIGO-Hanford and Virgo detectors. For LIGO Livingston, we use a minimum frequency of 30Hz~\cite{LIGOScientific:2020zkf}.

To generate gravitational waveforms in GR, we use four different waveform models: \texttt{IMRPhenomXPHM}~\cite{Pratten:2020ceb} (a state-of-art precessing spin multipolar gravitational waveform model), \texttt{IMRPhenomXHM}~\cite{Pratten:2020fqn,Garcia-Quiros:2020qpx} (a state-of-art multipolar gravitational waveform model for aligned spin binaries), \texttt{SEOBNRv4\_ROM}~\cite{Cotesta:2020qhw} (a reduced order based effective-one-body model for aligned spin binaries) and \texttt{IMRPhenomHM}~\cite{London:2017bcn} (first multipolar gravitational waveform model for aligned spin binaries). While we use
\texttt{IMRPhenomXPHM} as our \textit{default} choice of waveform model, additional models have been used to understand waveform systematic (if any). In all cases, waveforms have been computed using \texttt{LALSuite} software library~\cite{lalsuite}.
To generate waveforms with generic deviations from GR, we then modify \texttt{IMRPhenomXPHM}, \texttt{IMRPhenomXHM}, \texttt{IMRPhenomHM} and \texttt{SEONNRv4HM\_ROM} waveforms following Eq.(\ref{eq:test}).
For comparison, we also employ \texttt{IMRPhenomPv3HM} \cite{khan2020including} (another multipolar gravitational waveform model for precessing spin binaries) as this model has been widely used to estimate the source properties of all LVC events.

\begin{figure}[t]
	\includegraphics[scale=0.5]{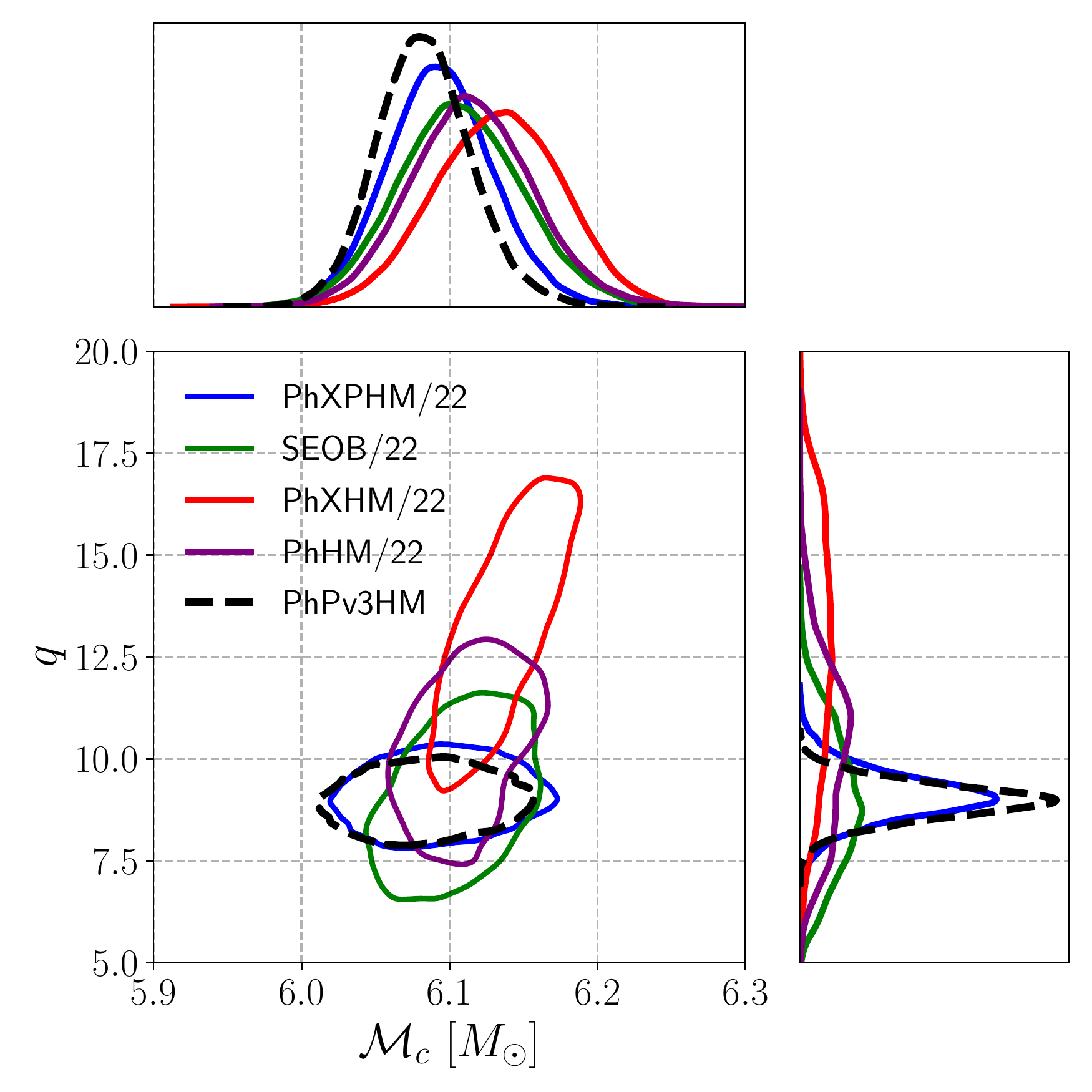}
	\caption{Comparison of chirp mass $\mathcal{M}_c$ and mass ratio $q$ posteriors inferred from the $(\ell,m)=(2,\pm2)$ modes of the GW190814 strain data using four different waveform models:  \texttt{IMRPhenomXPHM} (blue; {labeled as} \texttt{PhXPHM}), \texttt{IMRPhenomXHM} (red; {labeled as} \texttt{PhXHM}), \texttt{IMRPhenomHM} (magenta; {labeled as} \texttt{PhHM}) and \texttt{SEONNRv4HM\_ROM} (green; {labeled as} \texttt{SEOB}).  We show the estimated two-dimensional contours for 90\% confidence interval (middle panel) and one-dimensional kernel density estimates (KDEs) using Gaussian kernel (side panels). For comparison, we also show the estimates using all available modes in \texttt{IMRPhenomPv3HM} (black dashed line; {labeled as} \texttt{PhPv3HM}) model. Details are in Section \ref{Sec:constrains}.
	}
	\label{Fig:mc_q}
\end{figure}

\begin{figure}[t]
	\includegraphics[scale=0.5]{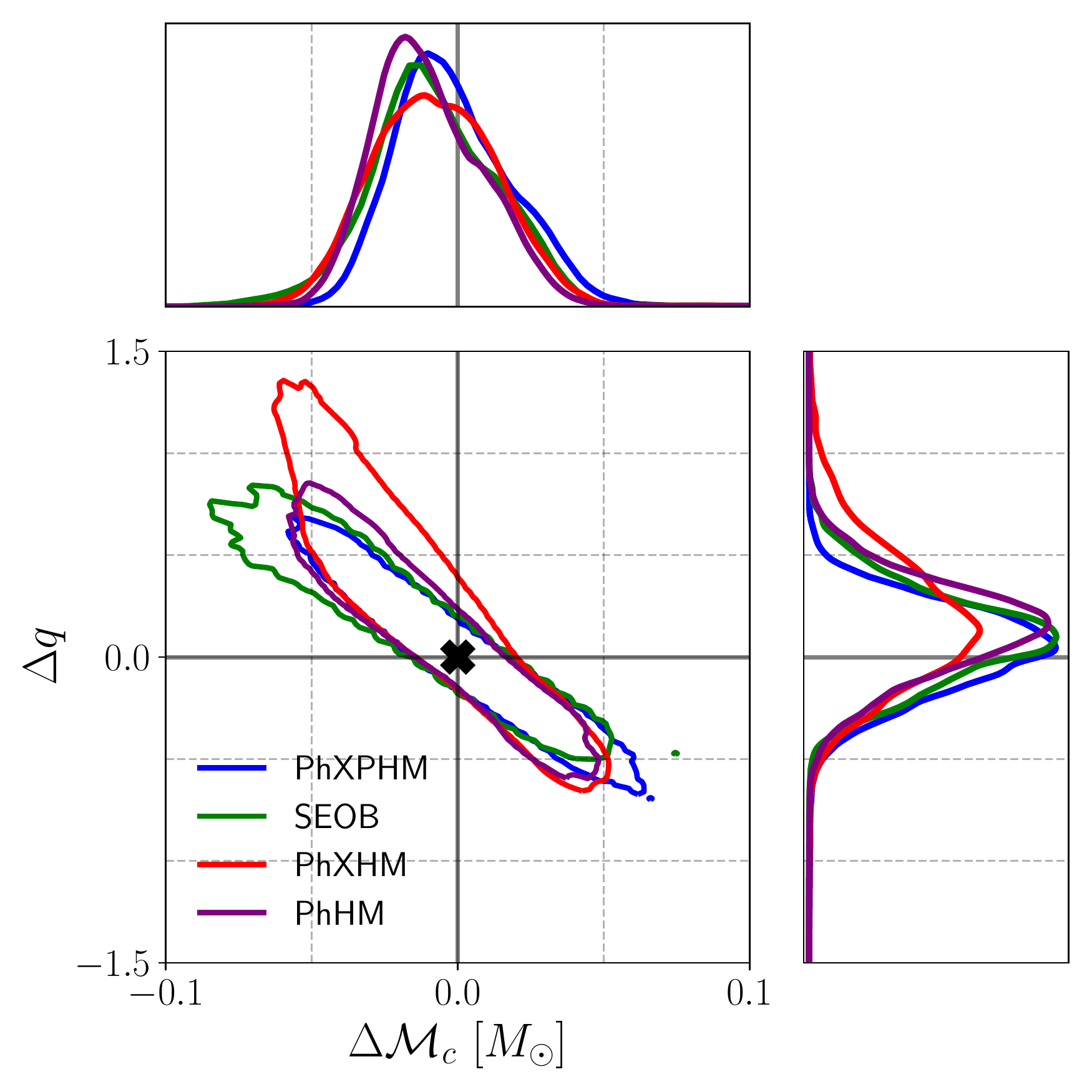}
	\caption{Comparison of the posteriors of the deviation parameters $\Delta \mathcal{M}_c$ and $\Delta q$ inferred from GW190814 strain data using four different waveform models: \texttt{IMRPhenomXPHM} (blue; {labeled as} \texttt{PhXPHM}), \texttt{IMRPhenomXHM} (red; {labeled as} \texttt{PhXHM}), \texttt{IMRPhenomHM} (magenta; {labeled as} \texttt{PhHM}) and \texttt{SEONNRv4HM\_ROM} (green; {labeled as} \texttt{SEOB}). We show the estimated two-dimensional contours for 90\% confidence interval (middle panel) and one-dimensional kernel density estimates (KDEs) using Gaussian kernel (side panels).
	The posteriors are fully consistent with the GR prediction of $\Delta \mathcal{M}_c = \Delta q = 0$ (shown by a ``+'' sign in the center panel and by thin black lines in all panels).
	Details are in Section \ref{Sec:constrains}.
	}
	\label{Fig:delta_mc_delta_q}
\end{figure}

\subsection{Choice of priors}
Our assumptions for the prior PDFs are mostly identical to the LVC analysis of GW190814~\cite{LIGOScientific:2020zkf}.
\begin{itemize}
	\item[i)] We choose uniform priors for the chirp mass and mass ratio ($5 M_\odot \le \mathcal{M}_{c} \le 100 M_\odot$ and $4 <= q <=20$), as defined in the rest frame of the Earth. The higher value ($q=4$) chosen for the lower limit of mass ratio prior does not affect our result as the resulting $q$ posterior does not have any support for values $q<4$. This is done in order for us to use a wider prior for the deviation parameters $\Delta \mathcal{M}_{c}$ and $\Delta q$.
	\item[ii)] Uniform priors are also used for the component dimensionless spins ($0.0 \le \chi_1\le 1.0$ and $0.0 \le \chi_2\le 1.0$), with spin-orientations taken as uniform on the unit sphere.
	\item [iii)] The prior on the luminosity distance is taken such that it is uniform in co-moving volume~\cite{ashton2019bilby} with  $1$ Mpc $\leq D_L \leq 2000$ Mpc.
	\item [iv)] For the orbital inclination angle $\theta_{JN}$, we assume a uniform prior over $0 \leq \cos\theta_{JN} \leq 1$.
	\item [v)] Priors on the sky location parameters $\alpha,\delta$ (right ascension and declination) are assumed to be uniform over the sky with periodic boundary conditions.
	\item [vi)] Finally, for the deviation parameters: we use uniform priors over $ -3M_\odot \le \Delta \mathcal{M}_{c} \le 3M_\odot$ and $-3 \le \Delta q \le 3$.
\end{itemize}

%==========================================================================
%==========================================================================
%==========================================================================
\section{Results} 
\label{Sec:results}
We now investigate how well the deviation parameters can be measured from GW190814 strain data and discuss the implication of our findings on the validity of `no-hair' theorem for binary black holes in GR. Posteriors of select parameters of interest are also shown.
%==========================================================================
%==========================================================================
%==========================================================================
\subsection{Constraining the deviation parameter} 
\label{Sec:constrains}
We perform the higher modes consistency tests on the detected GW190814 strain data using four different waveform models: \texttt{IMRPhenomXPHM}, \texttt{IMRPhenomXHM}, \texttt{IMRPhenomHM} and \texttt{SEONNRv4HM\_ROM}. While \texttt{IMRPhenomXPHM} is a precessing spin model, all remaining models are restricted to aligned-spin systems. The published LVC analysis~\cite{LIGOScientific:2020zkf}, however, implies that GW190814 is a low-spinning system with effective inspiral spin $\chi_{\rm eff}=-0.001^{+0.060}_{-0.061}$ and an effective precession parameter $\chi_p \le 0.07$. The aligned-spin models are, therefore, expected to capture the binary properties reasonably well. Employing different waveform models with varying modelling strategies will help us to identify any waveform systematic (if any) that may affect the accuracy of the higher modes consistency test particularly in the context of GW190814.

Higher modes consistency test simultaneously estimates the chirp mass $\mathcal{M}_c$ and mass ratio $q$ from the dominant $(\ell, m) = (2,\pm2)$ modes and the deviation parameters $\Delta \mathcal{M}_c$ and $\Delta q$ from the available higher order modes along with other parameters. Fig. \ref{Fig:mc_q} shows the inferred chirp mass $\mathcal{M}_c$ and mass ratio $q$ as estimated from the $(2,\pm2)$ modes using various waveform models. We show both the marginalized one-dimensional kernel density estimates (KDEs) using Gaussian kernel (side panels) and two-dimensional contours for 90\% confidence intervals (middle panel). For comparison, we also show the inferred values estimated using \textit{all} available modes in \texttt{IMRPhenomPv3HM} model (black dashed lines). We note that the $(\ell,m)=(2,\pm2)$ mode \texttt{IMRPhenomXPHM} estimates of $\mathcal{M}_c$ and $q$ matches \texttt{IMRPhenomPv3HM} estimates using \textit{all} available modes. Aligned spin models, on the other hand, yield broader posteriors (and contours). However, these estimates do overlap with each other implying that the broadening of posteriors is due to the loss of information as these models assume the binaries to have aligned spin components only.

Constraints on the deviation parameters $\Delta \mathcal{M}_c$ and $\Delta q$ in the higher order modes (obtained using different waveform models) are shown in Fig.~\ref{Fig:delta_mc_delta_q}. We find that the deviation parameters are consistent with the GR values: a zero deviation is within the 90\% credible interval across waveform models. It is interesting to note that, even though the $(\ell,m)=(2,\pm2)$ mode estimates of $\mathcal{M}_c$ and $q$ obtained from different waveform models (mostly between the estimates inferred with a precessing model and aligned spin models) have some noticeable differences, the posteriors for deviation parameters agree quite well to each other. The only exception is \texttt{IMRPhenomXHM} which prefers a longer tail for $\Delta q$ (and for $q$ too). In all cases, the fractional deviation  $\frac{\Delta \mathcal{M}_c}{\mathcal{M}_c} (\%)$ and $\frac{\Delta q}{q} (\%)$ are only a few percent.

This exercise provides interesting insights into the waveform systematic. Various waveform models employed here include different higher order harmonics (see Table \ref{Tab:modes}). However, no significant biases in $\Delta \mathcal{M}_c$ and $\Delta q$ across waveform models (except \texttt{IMRPhenomXHM}) implies that, even though there are differences in  $\mathcal{M}_c$ and $q$ estimates across different models, inferred values from the $(2,\pm2)$ and higher order modes for a particular waveform model is consistent with each other.

% ------------------------------------------------------------------------------------------------------
% ------------------------------------------------------------------------------------------------------
\begin{table}
	\centering
	\begin{tabular}{p{0.15\textwidth}| p{0.3\textwidth}}
		\toprule
		Waveform model                  & Available HOM         \\
		\hline
		{\texttt{IMRPhenomXPHM}} & $(2,\pm1),(3,\pm2),(3,\pm3),(4,\pm4)$ \\
		\hline
		{\texttt{IMRPhenomXHM}} & $(2,\pm1),(3,\pm2),(3,\pm3),(4,\pm4)$ \\
		\hline
		{\texttt{IMRPhenomHM}} & $(2,\pm1),(3,\pm2),(3,\pm3),(4,\pm3),(4,\pm4)$ \\
		\hline
		{\texttt{SEOBNRv4\_ROM}} & $(2,\pm1),(3,\pm3),(4,\pm4),(5,\pm5)$ \\
		\botrule	
	\end{tabular}
	%}
	\caption{Available higher order modes i.e. modes apart from the dominant $(\ell,m)=(2,\pm2)$ mode in various waveform models used in this study.}
	\label{Tab:modes}
\end{table}
% ------------------------------------------------------------------------------------------------------
% ------------------------------------------------------------------------------------------------------

\begin{figure}[t]
	\includegraphics[scale=0.57]{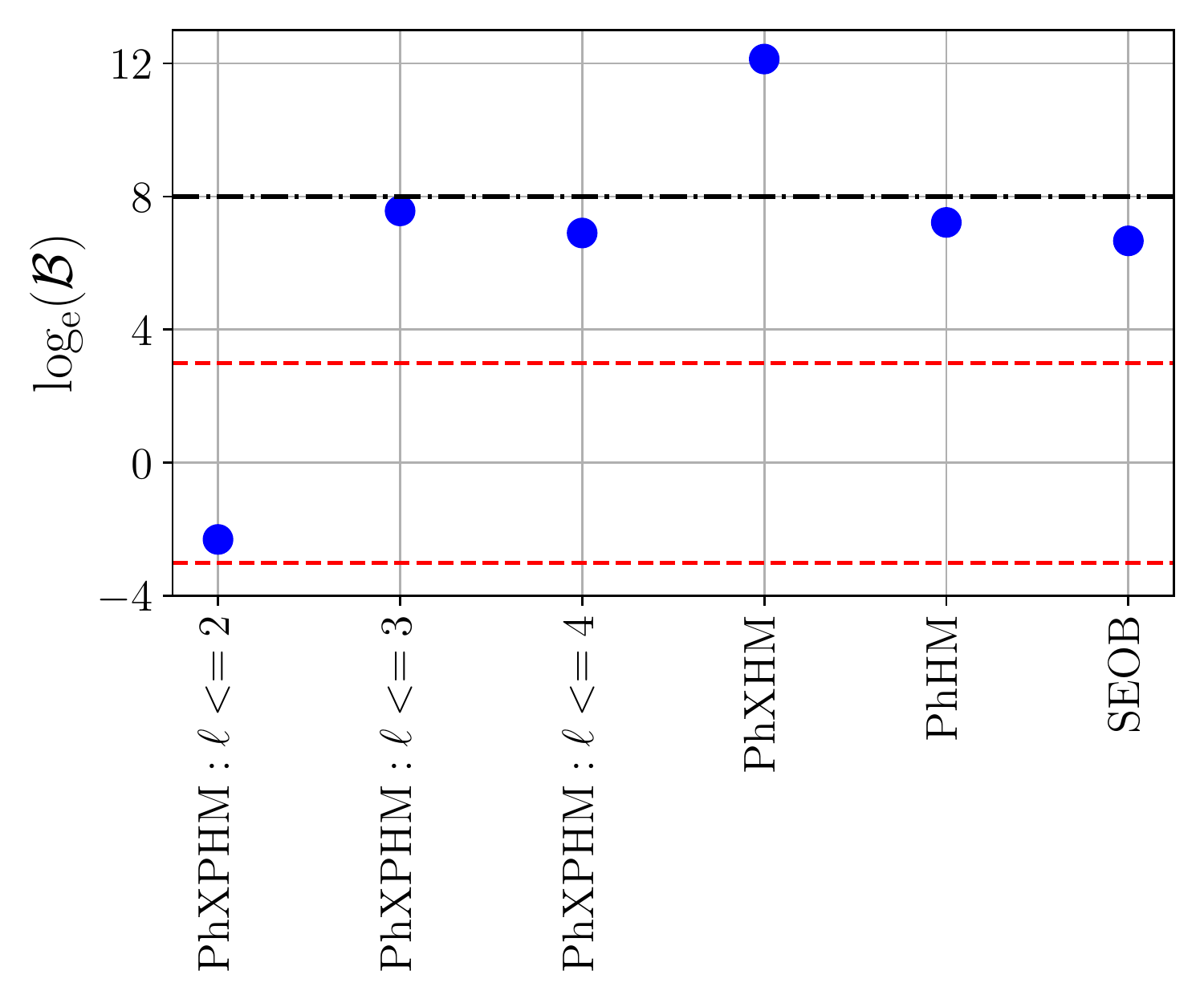}
	\caption{$\rm log_{e}(\mathcal{B})$ in favor of the hypothesis that the observed strain data is consistent with the multipolar structure expected for signals emitted from binary black holes in GR. $\rm log_{e}(\mathcal{B})=0$ indicates both the GR and deviation from binary in GR hypothesis are equally likely. To put things into perspective, we also show $\rm log_{e}(\mathcal{B})=\pm3$ (which provides enough hint for a hypothesis) and  $\rm log_{e}(\mathcal{B})=\pm8$ (which strongly favors a hypothesis).
	Details are in Section \ref{Sec:testing_nohair}.
	}
	\label{Fig:bayes_factor}
\end{figure}

%==========================================================================
%==========================================================================
%==========================================================================
\subsection{Testing the `no-hair' theorem} 
\label{Sec:testing_nohair}
Next, we test the `no-hair' hypothesis and quantify its evidence in data. This is done in two steps. First, we analyze the data assuming a GR hypothesis: this involves estimating the set of 15 parameters in GR that describes the signal. Then, we repeat the analysis with a non-GR hypothesis where the signal is modified according to Eq.(\ref{eq:polarizations_modgr}) to allow possible departure from the multi-polar structure of gravitational radiation in GR. We then compute the Bayes factor in favor of the `no-hair' theorem. In Fig. \ref{Fig:bayes_factor}, we show the estimated Bayes factors (in logarithmic scale) for the 'no-hair' hypothesis in GR estimated using different waveform models. To put things into perspective, a $\rm log_{e}(\mathcal{B})=0$ indicates that both the `no-hair' hypothesis in GR and the hypothesis supporting a deviation from it are equally likely while $\rm log_{e}(\mathcal{B})=\pm3$ ( $\rm log_{e}(\mathcal{B})=\pm8$) implies some support for a particular hypothesis (implies that the data strongly favors a particular hypothesis). For all of the waveform models, we find that the data has overwhelming support the `no-hair' hypothesis with $\rm log_{e}(\mathcal{B}) \sim 7-8$. Therefore, our study suggests that the data favors the GR hypothesis where this deviation parameters are set to zero.

\begin{figure}[t]
	\includegraphics[scale=0.5]{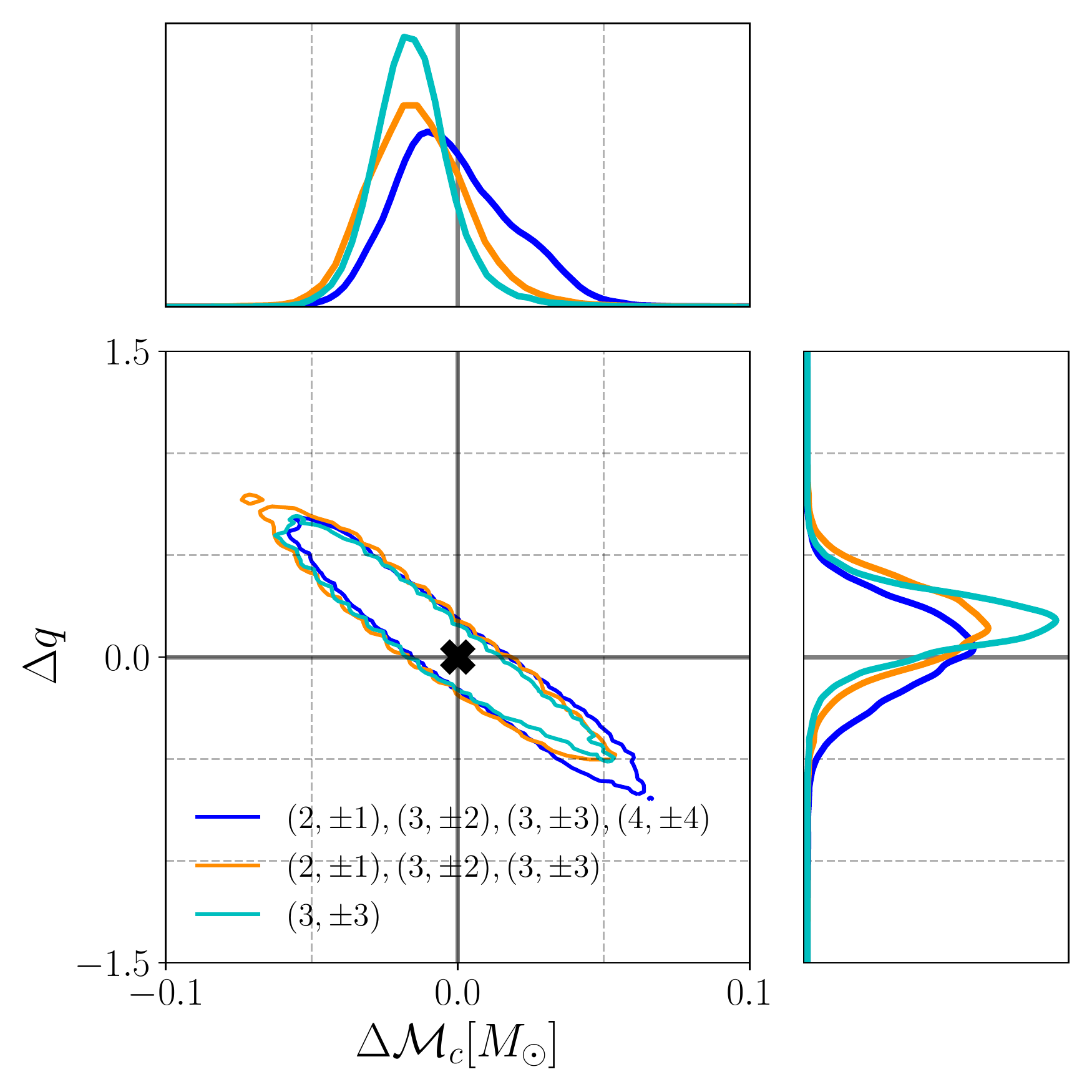}
	\caption{Comparison of the posteriors of the deviation parameters $\Delta \mathcal{M}_c$ and $\Delta q$ inferred from GW190814 strain data using different mode content in the higher order modes:  (i) $(\ell,m)=\{(2,\pm1),(3,\pm2),(3,\pm3),(4,\pm4)\}$ (blue), (ii) $(\ell,m)=\{(2,\pm1),(3,\pm2),(3,\pm3)\}$ (orange) and (iii) $(\ell,m)=\{(3,\pm3)\}$ (cyan).  We show the estimated two-dimensional contours for 90\% confidence intervals (middle panel) and one-dimensional kernel density estimates (KDEs) using Gaussian kernel (side panels).
	The posteriors are fully consistent with the GR prediction of $\Delta \mathcal{M}_c = \Delta q = 0$ (shown by a ``+'' sign in the center panel and by thin black lines in all panels).
	Details are in Section \ref{Sec:effect_of_hm}.
	}
	\label{Fig:effects_of_hm}
\end{figure}

%==========================================================================
%==========================================================================
%==========================================================================
\subsection{Effects of higher order modes} 
\label{Sec:effect_of_hm}
We now investigate how the mode contents in the higher order harmonics affect the test accuracy. We demonstrate the effect of higher order modes using our default choice of waveform model \texttt{IMRPhenomXPHM}. To do this, we perform the higher modes consistency test with different combination of modes:
\begin{itemize}
	\item  First, we only include the $(\ell,m)=(2,\pm1)$ modes in our higher harmonics.
	\item More modes are then gradually included. We repeat the exercise with another two configurations of mode content: (i) $\{(2,\pm1),(3,\pm2),(3,\pm3)\}$ (including all available $\ell \le 3$ modes in \texttt{IMRPhenomXPHM}); and (ii) $\{(2,\pm1),(3,\pm2),(3,\pm3),(4,\pm4)\}$ (including all available $\ell \le 4$ modes in \texttt{IMRPhenomXPHM}).
	\item Finally, we note that the LVC analysis finds strong support for the $(\ell,m)=(3,\pm3)$ modes in the strain data apart from the dominant $(\ell,m)=(2,\pm2)$ modes~\cite{LIGOScientific:2020zkf}. We, therefore, redo our analysis for the case where the higher harmonics only include the $(\ell,m)=(3,\pm3)$ mode.
\end{itemize}
Fig. \ref{Fig:bayes_factor} shows the recovered Bayes factors in support of the GR hypothesis for varying mode content. Interestingly, we observe that for the case when \textit{only} $(\ell,m)=(2,\pm1)$ modes are included in the higher harmonics, the deviation parameters remains mostly unresolved (not shown) and the test favors ($\rm log_{e}(\mathcal{B}) \sim -3.5$) the hypothesis that the data is better described by a departure from the multi-polar structure in radiation as expected in GR.
However, as we include other important modes, specially $(\ell,m)=(3,\pm3)$ modes, $\rm log_{e}(\mathcal{B})$ becomes positive again implying increased support for the GR hypothesis. In Fig. \ref{Fig:effects_of_hm}, we show the posteriors for the deviation parameters $\Delta \mathcal{M}_c$ and $\Delta q$. We notice that even though the posteriors are all consistent with GR expectation of zero, they show an interesting trend. Both $\Delta \mathcal{M}_c$ and $\Delta q$ has sharper peaks when they are estimated from only the $(\ell,m)=(3,\pm3)$ modes. Furthermore, the peaks appear to be shifted from zero even though the posteriors are still consistent with zero at 90\% credible intervals. As we include more and more modes, the posteriors becomes broader but the peaks shift towards zero. This study clearly shows the need to have as many higher order modes as possible so that not only our usual PE estimates are correct, but also any tests of GR, specially higher modes consistency tests or quasi-normal modes consistency tests, are able to provide trustful results. 

%==========================================================================
%==========================================================================
%==========================================================================
\subsection{Nature of the compact objects} 
To test our null hypothesis that GW190814 signal is produced by a BBH in GR, we constrain the deviation parameters $\Delta \mathcal{M}_c$ and $\Delta q$.
For BBH signal in GR, these deviation parameters are expected to be consistent with zero.
Non-zero values $\Delta \mathcal{M}_c$ and $\Delta q$ would imply a possible departure from a BBH in GR. 
Such departures would point to either (i) a violation of GR; or (ii) a merger of compact objects that are not black holes such as neutron stars or other exotic objects; or (iii) the possibility that the signal might have been modified due to other astrophysical processes (such as environmental effects or lensing) that have been neglected in analysis. 
We point out that the higher modes consistency test can distinguish a NSBH signal from a BBH when the signal-to-noise ratio is high (cf. Fig. 11 of Ref.~\cite{Islam:2019dmk}). 
Overall, our results in Section \ref{Sec:testing_nohair} favors the hypothesis that the GW signal from GW190814 is consistent with the multipolar structure of radiation expected from a binary black hole merger in GR over possible departure from it by $\rm log_{e}(\mathcal{B}) \sim 7-8$. 
However, we must caution that it is still possible that the signal have imprints of beyond GR effects or signature of compact objects apart from BBH; and those imprints may be small enough that they have not been captured by this test in the current signal-to-noise ratio. 

\subsection{Implications for future detectors with improved sensitivity} 
\label{Sec:rescaled_psd}

Our findings in Section \ref{Sec:effect_of_hm} suggests that any higher modes consistency tests using only select higher order harmonics may falsely identify a GR signal if the signal has high SNRs. Such scenarios may occur as our detectors gradually become more sensitive. We investigate this possibility using a simplistic approach that fits in between a real GW data analysis and a simulated data analysis. We decide to use the actual strain and calibration data for GW190814 but re-scale the PSD by some factor to simulate the same event but in a more sensitive network of detectors. We consider two different cases where the PSDs across detectors have been re-scaled by a factor of $\frac{1}{7.5}$ and $\frac{1}{12.5}$ apart from the case where we use actual PSDs. We then perform the higher modes consistency tests employing \texttt{IMRPhenomXPHM} model. Furthermore, we use only $(\ell,m)=(3,\pm3)$ modes in the higher harmonics. While we could have included all the available modes in the model, our choice is motivated by the findings that when only $(\ell,m)=(3,\pm3)$ modes are included in the higher harmonics, both $\Delta \mathcal{M}_c$ and $\Delta q$ have sharper peaks that are slightly offset from zeros (Fig. \ref{Fig:effects_of_hm}). We find that, due to the improvement in sensitivity, $\Delta \mathcal{M}_c$ and $\Delta q$ posteriors become sharper as expected when the PSDs are scaled. Interestingly, for the case where the PSDs have been scaled by a factor of $\frac{1}{12.5}$, higher modes consistency test yields $\Delta \mathcal{M}_c$ and $\Delta q$ posteriors that are inconsistent with GR. 
We must mention that the exercise of rescaling the PSDs may exaggerate the effect of the noise that is on top of the signal and may bias the parameter estimation.
Nonetheless, this demonstrates that future models need to include more higher order modes to avoid scenarios where a purely GR signal may be labeled as a violation of GR as a result of the loss of information due to the unavailability of many higher order harmonics. 

\begin{figure}[t]
	\includegraphics[scale=0.5]{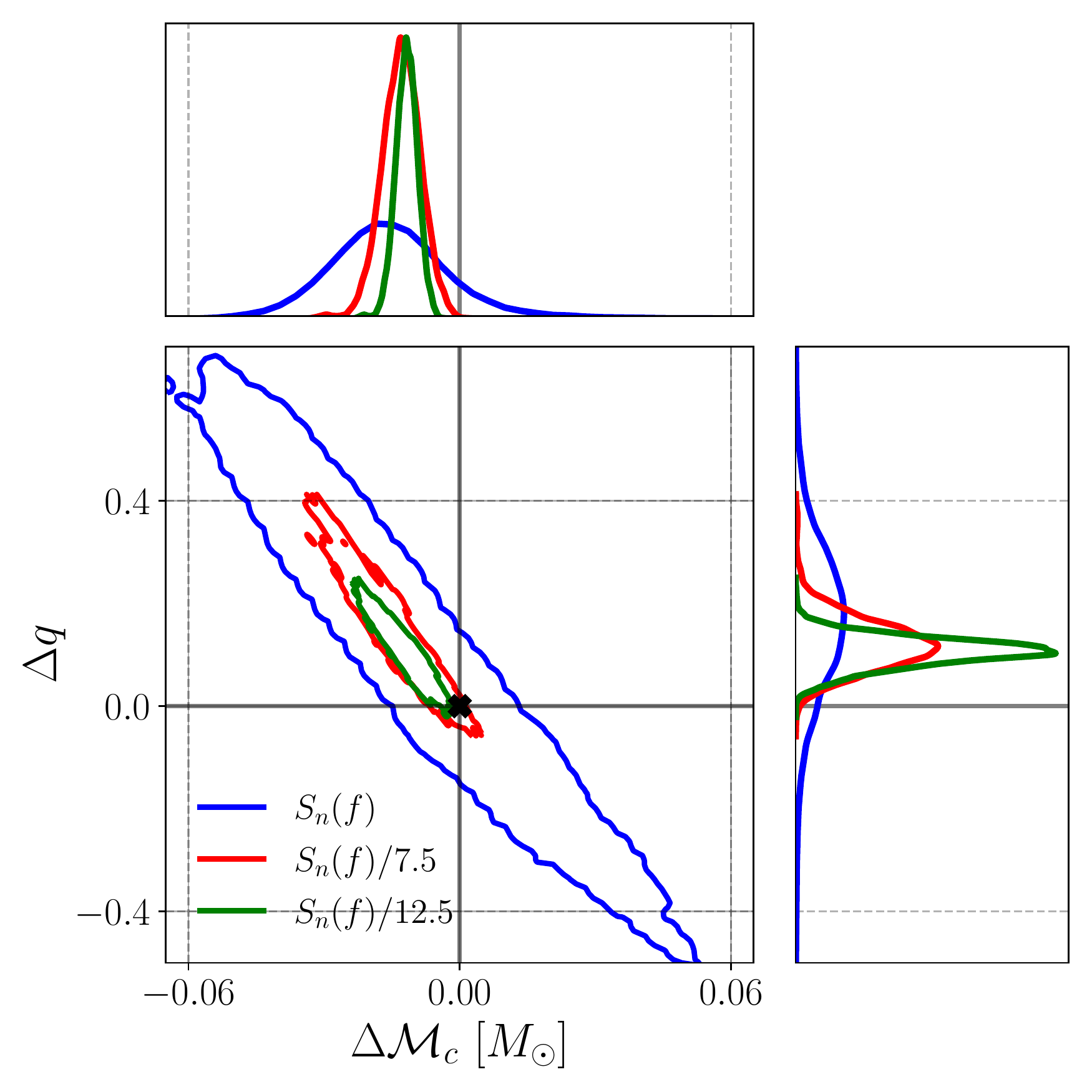}
	\caption{Comparison of the posteriors of the deviation parameters $\Delta \mathcal{M}_c$ and $\Delta q$ inferred from GW190814 strain data employing \texttt{IMRPhenomXPHM} model (only the $(\ell,m)=(3,\pm3)$ modes are included in the higher harmonics) with rescaled PSDs for all three detectors to imitate a future network of detectors with improved sensitivity. Apart from the results obtained using the actual PSDs (blue), we show results inferred from two other cases where the PSDs across detectors have been scaled by a factor of $1/7.5$ (red) and $1/12.5$ (green).
	We show the estimated two-dimensional contours for 90\% confidence intervals (middle panel) and one-dimensional kernel density estimates (KDEs) using Gaussian kernel (side panels).
	GR prediction of $\Delta \mathcal{M}_c = \Delta q = 0$ is shown by a ``+'' sign in the center panel and by thin black lines in all panels.
	When the PSDs are rescaled by a factor of $1/7.5$, resultant posteriors become inconsistent with GR.
	Details are in Section \ref{Sec:rescaled_psd}.
	}
	\label{Fig:rescaled_psd}
\end{figure}

%==========================================================================
%==========================================================================
%==========================================================================
\section{Final remarks} 
\label{Sec:conclusion}

We provide a comprehensive analysis of the evidence for `no-hair' theorem for BBH in GR using GW190814 strain data. To test the validity of the `no-hair' theorem in GR, we perform `higher modes consistency' test proposed in~Refs. \cite{Islam:2019dmk,Dhanpal:2018ufk}. The test checks whether the estimation of chirp mass $\mathcal{M}_c$ and mass ratio $q$ from the dominant quadrupolar $(\ell,m)=(2,\pm2)$ modes are consistent with values inferred from the higher order harmonics. This is done by introducing additional deviation parameters in masses for the higher order modes:  $\Delta \mathcal{M}_c$ and $\Delta q$. If the signal is originated from a binary black hole system as described in GR, these two deviation parameters will be zero. A non-zero value will point to possible modifications of gravity or astrophysical phenomenon not considered in analysis or even exotic nature of the compact objects. To detect (or nullify) any possible biases in inferred $\Delta \mathcal{M}_c$ and $\Delta q$ parameters originating from the choice of waveform models, we perform the test using four different models having varying mode contents and accuracy: \texttt{IMRPhenomXPHM}, \texttt{IMRPhenomXHM}, \texttt{IMRPhenomHM} and \texttt{SEONNRv4HM\_ROM}. These models also have different domain of validity. While \texttt{IMRPhenomXPHM} models GW signal from generically spinning binaries, all other models are restricted to only aligned-spin binary black holes. As the detected binary has negligible spin, these models are expected to describe the system reasonably well. 

\begin{figure}[t]
	\includegraphics[scale=0.5]{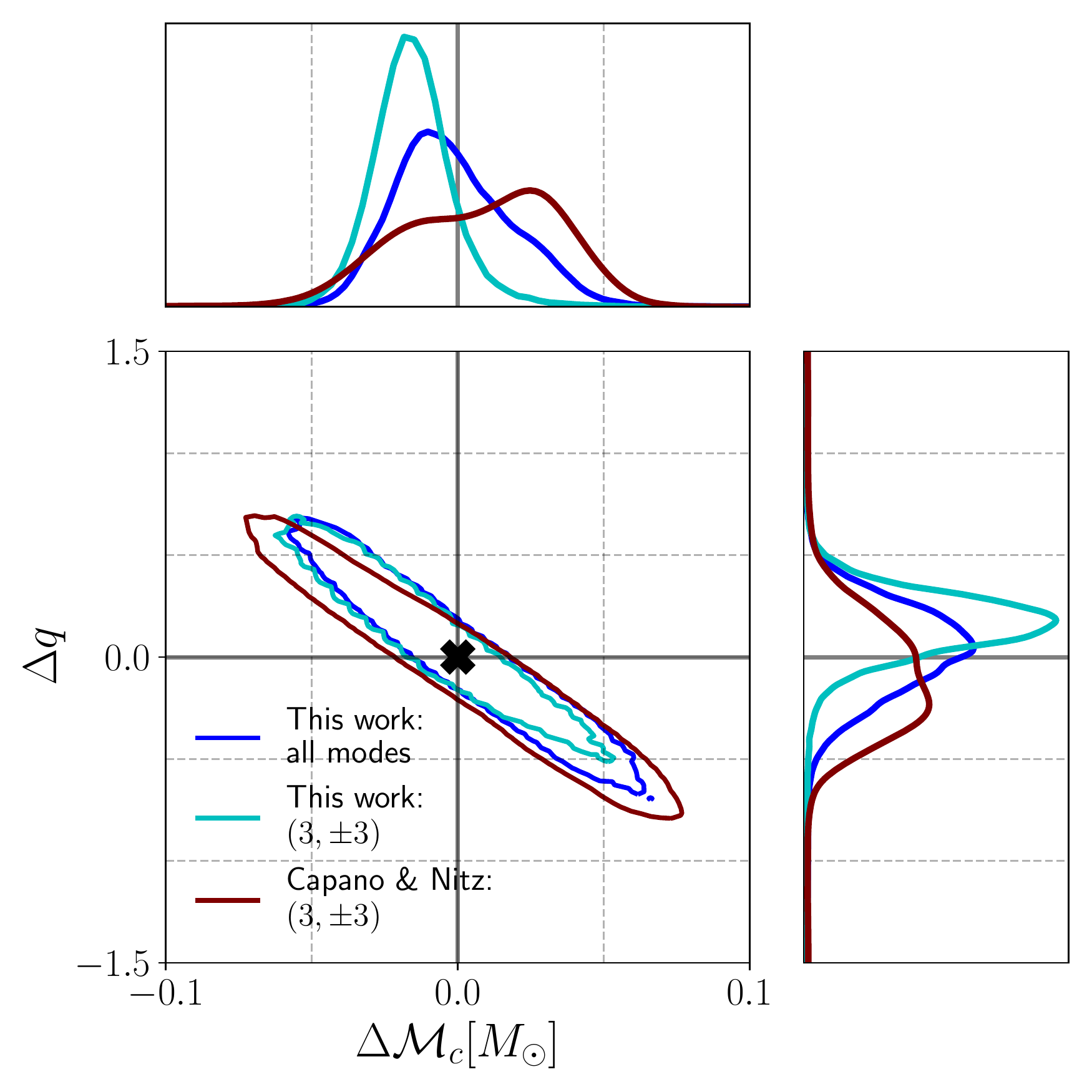}
	\caption{Comparison of the posteriors of the deviation parameters $\Delta \mathcal{M}_c$ and $\Delta q$ inferred using our
		higher modes consistency test~\cite{Islam:2019dmk,Dhanpal:2018ufk} and results obtained in Ref.~\cite{Capano:2020dix}.
		We show our $\Delta \mathcal{M}_c$ and $\Delta q$ estimates using  (i) all available higher order modes in the waveform model (blue) and (ii) using only $(\ell,m)=\{(3,\pm3)\}$ mode in the higher harmonics (cyan).  
		Results obtained in Ref~\cite{Capano:2020dix} using only the $(\ell,m)=\{(3,\pm3)\}$ mode in the higher harmonics through a different version of the higher modes consistency test is shown in brown.
		We show the estimated two-dimensional contours for 90\% confidence intervals (middle panel) and one-dimensional kernel density estimates (KDEs) using Gaussian kernel (side panels).
		The posteriors are fully consistent with the GR prediction of $\Delta \mathcal{M}_c = \Delta q = 0$ (shown by a ``+'' sign in the center panel and by thin black lines in all panels).
		Details are in Section \ref{Sec:conclusion}.
	}
	\label{Fig:compare_capano}
\end{figure}

We find that the deviation parameters $\Delta \mathcal{M}_c$ and $\Delta q$ are broadly consistent with zeros irrespective of the waveform model used (Fig. \ref{Fig:delta_mc_delta_q}). We note that similar test have recently been applied to GW190412 and GW190814 strain data~\cite{Capano:2020dix}. The test includes only $(2,\pm2)$ and $(3,\pm3)$ modes in their study and considers
two different ways to look for the consistency: (i) by allowing all intrinsic parameters including spins to assume different values in different modes; and (ii) then allowing only the masses to have different values across modes. In both the cases, orbital phases are free to have different values in $(2,\pm2)$ and $(3,\pm3)$ modes. In Fig. \ref{Fig:compare_capano}, we compare results obtained using the later strategy in Ref.~\cite{Capano:2020dix} against that of ours obtained using (i) all available higher order modes in the waveform model and (ii) using only $(\ell,m)=\{(3,\pm3)\}$ mode in the higher harmonics. We find that while $\Delta \mathcal{M}_c$ and $\Delta q$ posteriors have visible differences, all three posteriors are consistent with GR expectation and the fractional deviations $\frac{\Delta \mathcal{M}_c}{\mathcal{M}_c} (\%)$ and $\frac{\Delta q}{q} (\%)$ are only a few percent in all cases. We attribute the remaining differences to the differences in the formulation of the consistency tests. For example, we do not allow the orbital phase to take different values in different modes.

GW190814 data overwhelmingly supports the hypothesis that the signal obeys the `no-hair theorem' in GR (i.e. the signal is consistent with the multipolar structure of radiation as expected from a merging binary black hole in GR) over hypothesis proposing a deviation from the `no-hair theorem' (Fig.\ref{Fig:bayes_factor}). The log Bayes factor computed in favor of the `no-hair theorem' is found to be $\rm log_{e}(\mathcal{B})\sim7-8$. This implies that the binary is likely to be a binary black hole in GR. We must caution that it is, however, possible that the a binary is not a BBH or the signal actually obeys a theory of gravity different from GR but the test is unable to pick up those signatures at the current signal-to-noise. 

We further show that including several higher order modes are crucial for consistency tests like this one. We observe that when the higher harmonics only include the $(\ell,m)=(2,\pm1)$ modes, the test favors a violation of GR even though the deviation parameters are mostly unresolved (Fig. \ref{Fig:bayes_factor}). This is due to the inability of the waveform model to account for neglected higher order modes likes $(\ell,m)=(3,\pm3)$ modes. Similar results have previously been obtained for cases where the higher-order modes are present in the simulated GR signal but the recovery template used to test GR includes only dominant $(2,\pm2)$ modes~\cite{Pang:2018hjb}. Next, we also show that while having only $(\ell,m)=(3,\pm3)$ modes in the higher harmonics would yield results consistent with GR expectation for current detectors (Fig. \ref{Fig:effects_of_hm}), it might not be enough when detector sensitivity improves in future. We demonstrate this by rescaling the PSDs by some constant factor and redoing the analysis. We find that when the sensitivity improves by at least a factor of $\sim12.5$, this test triggers a false alarm if none of the higher harmonics apart from $(\ell,m)=(3,\pm3)$ modes are included (Fig. \ref{Fig:rescaled_psd}). 

Apart from providing support for the `no-hair theorem', our study therefore calls for having as many higher order modes as possible in anticipation of future improvements in GW detectors.
Our study employs four different state-of-art waveform models from only the Phenom and EOB families. As none of the existing NR surrogate waveform model covers mass ratio $q\sim10$ for spinning binaries, we have not been able to use it in our analysis. Works are, however, underway to build NR-based or NR-tuned perturbation theory based surrogate waveform models for high mass binaries with spins. Availability of such models having more spherical harmonic modes compared to the existing phenomenological and EOB models may further improve our ability to test the `no-hair' theorem with detected GW signals. We leave that for future explorations.\\

%==========================================================================
%==========================================================================
%==========================================================================
 
\begin{acknowledgments}
T.I. thanks Vijay Varma, Scott Field, Gaurav Khanna, Max Isi, Harald Pfeiffer and Juan Calderon Bustillo for helpful feedback on the manuscript; Parameswaran Ajith, Ajit Kumar Mehta and Abhirup Ghosh for numerous discussion during the development of the test (cf. \cite{Islam:2019dmk}); Carl-Johan Haster for some helps related to plotting and Rory Smith for his comments on Bayes factor computation.
Simulations were performed on CARNiE at the Center for Scientific Computing and Visualization Research (CSCVR) of UMassD, which is supported by the ONR/DURIP Grant No. N00014181255 and the MIT Lincoln Labs {\em SuperCloud} GPU supercomputer supported by the Massachusetts Green High Performance Computing Center (MGHPCC). 
This research was supported in part by the Heising-Simons Foundation, the Simons Foundation, and National Science Foundation Grant No. NSF PHY-1748958.
The author acknowledge support of NSF Grants No. PHY-1806665 and No. DMS-1912716.

This research has made use of data, software and/or web tools obtained from the Gravitational Wave Open Science Center (https://www.gw-openscience.org/ )~\cite{LIGOScientific:2019lzm}, a service of LIGO Laboratory, the LIGO Scientific Collaboration and the Virgo Collaboration. LIGO Laboratory and Advanced LIGO are funded by the United States National Science Foundation (NSF) as well as the Science and Technology Facilities Council (STFC) of the United Kingdom, the Max-Planck-Society (MPS), and the State of Niedersachsen/Germany for support of the construction of Advanced LIGO and construction and operation of the GEO600 detector. Additional support for Advanced LIGO was provided by the Australian Research Council. Virgo is funded, through the European Gravitational Observatory (EGO), by the French Centre National de Recherche Scientifique (CNRS), the Italian Istituto Nazionale di Fisica Nucleare (INFN) and the Dutch Nikhef, with contributions by institutions from Belgium, Germany, Greece, Hungary, Ireland, Japan, Monaco, Poland, Portugal, Spain. This is LIGO Document Number DCC-P2100392.
\end{acknowledgments} 

%==========================================================================
%==========================================================================
%==========================================================================

%%%%%%%%%%%%%%%%%%%%%%%%%%%%%%%%%%%%%%%%%%%%%%%%%%%%%%%%%%%%%%%%%%%%%%%%%%%%%%%
\section*{References}
%%%%%%%%%%%%%%%%%%%%%%%%%%%%%%%%%%%%%%%%%%%%%%%%%%%%%%%%%%%%%%%%%%%%%%%%%%%%%%%
\bibliography{hm_tgr}

\end{document}